\documentclass[%
 reprint,
 amsmath,amssymb,
 aps,
]{revtex4-2}

\usepackage{graphicx}
\usepackage{dcolumn}
\usepackage{bm}

\begin{document}

\preprint{APS/123-QED}

\title{Semi Universal relation to understand matter properties at neutron star interiors}

\author{Ritam Mallick}
\email{mallick@iiserb.ac.in}
 \affiliation{Department of Physics,\\ Indian Institute of Science Education and Research Bhopal, Bhopal, India.}
\author{Debojoti Kuzur}%
 \email{debojoti16@iiserb.ac.in}
\affiliation{%
 Department of Physics,\\ Indian Institute of Science Education and Research Bhopal, Bhopal, India.
}%
\author{Rana Nandi}%
\email{nandi.rana@gmail.com}
\affiliation{%
	Department of Physics,Polba Mahavidhyalaya, Hoogly, West Bengal 712148, India.
}%

\date{\today}

\begin{abstract}
The occurrence of quark matter at the center of neutron stars is still in debate. This study defines some semi-empirical parameters that quantify the occurrence and the amount of quark matter at star interiors. These parameters show semi-universal relations across all the EoS. One parameter depends on the shifting of the keplerian mass-radius curve from the static one and shows it is a constant across all EoS. The Z-parameter shows how tidal deformability depends on the quark content of the star and the stiffness of the EoS. The quark content of the star also affects the compactness of the star, and its dependence is almost universal. The empirical parameter gives a bound on the quark content of the star and shows that if the amount of the quark content increases, the stars are likely to collapse into a black hole. It is seen that the change in the mass and radius after PT is linearly proportional to the mass of the parent NS. Given a hadronic EoS, bag constant, and quark coupling constant, one can have a critical mass of the neutron star and the maximum mass of the hybrid star for phase transition without any baryonic mass loss.
\end{abstract}

\maketitle


\section{Introduction}
The theory of strongly interacting matter, quantum chromodynamics (QCD), predicts hadrons to quarks and gluons deconfinement transition at high density and/or temperature \cite{shuryak}. The deconfinement transition at high temperature has been observed in heavy-ion collisions \cite{gyulassy,andronic} but the presence of quarks at high density remains unsolved. One of the naturally occurring laboratories of the dense matter is the cores of neutron stars (NS). However, the cores is not directly visible, and to have any information, we have to model NSs from the core to the surface and then match them with observations.

The matter properties at two extreme density limits at zero temperature are known with a certain degree of accuracy \cite{kruger}. At the low-density regime till nuclear saturation density the matter is in the hadronic phase and the modern nuclear theory (like chiral effective field theory) is quite accurate \cite{kukrela}. In the very high-density limit perturbative-QCD (pQCD) techniques with quarks and gluons as their degrees of freedom becomes reliable. This points to the fact that there is a deconfinement phase transition (PT) from hadrons to quarks happens at densities between these two limits. The cores of NSs at their heart bears these intermediate densities where PT can occur \cite{rezolla,prasad1,prasad2,ritam-mnras,oconnor}

In the last decade,  astrophysical observation has put severe constraints on the equation of state and has given hope of establishing the properties of matter at NS cores.
A big breakthrough came in the form of two very massive pulsars \cite{demorest,antonidis,cromartie}. However, mass measurement alone is not sufficient to eliminate any discrepancies in the EoS; subsequent accurate knowledge of NS radius is also needed. NS Interior Composition Explorer (NICER) was recently launched, which is estimated to measure the NS radius with $5\%$ accuracy. It recently measured the mass-radius of a pulsar PSR J0030+0451 (R=$12.71^{+1.14}_{-1.19}$ and M=$1.34^{+0.15}_{-0.16}$) \cite{riley,miller}

The detection of a gravitational wave (GW) from binary NS merger GW170817 came next \cite{abbott}. During the inspiral phase, both NSs 
induces a strong tidal deformation on the other due to their 
gravitational field \cite{hinderer}. The tidal deformability of an NS is related to its compactness (compactness $C=M/R$), and its information gets imprinted on the GW, which puts an additional constraint on the EoS \cite{annala,bauswein,most,rana,zhang}. GW170817 gave tidal deformability 
($\Lambda$) bound of $\Lambda \le 580$ \cite{abbott2}, which constrains the radius of a $1.4$ solar mass NS to be in the range of $12.9-13.5$ km. 

Even with new stringent constraints, NS core can shelter quark matter \cite{nandi-prc,annala}.  It would be advantageous to generate some general relation differing from each other depending on whether NS core has or does not have quark matter in them. 

\section{Formalism}
In order to define the structure of the NS we have used the Rotating Neutron Star (RNS) code \cite{rns}. 
In order to solve for the entire structure of the NS, the final equation that is needed is the EoS. For the EoS of the hadronic matter (HM) we consider relativistic 
mean-field (RMF) models: S271v2 (S27) \cite{Horowitz:2002mb}, BSR1,BRS2,BRS3,BRS4 \cite{BSR}, DD2 \cite{Typel:2009sy}, DDME \cite{DDME1} and a IOPBI \cite{IOPB}. 
At low density, the Baym-Pethick-Sutherland EoS \cite{Baym:1971pw} of the crust is added to all of these hadronic EoS in a thermodynamically consistent fashion. 
The EoS of the QM is constructed by adopting the modified MIT Bag model consisting of up ($u$), down ($d$) and strange ($s$) quarks and electrons \cite{chodos}. The strong interaction correction and the non-perturbative QCD effects are included via two effective parameters $a_4$ and $B_{\rm eff}$ \cite{alford,weissenborn}. 


The EoS is constructed to have nuclear matter at low density, mixed-phase (quarks and hadrons) at intermediate density range, and pure quark matter at high density. 
The $B_{\rm eff}$ and $a_4$ are chosen in such a way that the hybrid star (HS) formed by such mixed EoS satisfies all the present nuclear and astrophysical bounds.
The mixed-phase is constructed using the Gibbs construction. The hybrid EoS which at low density have HM at intermediate density has mixed-phase region (both hadrons and quarks) and at very high densities have pure quark region. If the mixed-phase region extends till very high densities, the NS constructed with these EoS will have a mixed-phase at their core, and pure quark matter will not appear in these stars. However, for some, the mixed-phase occurs at relatively lower densities; therefore, massive stars constructed with these EoS are likely to have a pure quark core followed by a mixed-phase region in the intermediate region and pure nuclear outer surface. In the rest of the article, our quark core implies a mixed phase quark core with $R^q$ ($R_e^q$ and $R_p^q$ for equatorial and polar radius respectively in case of the rotating star) can be calculated by numerically solving the change in density of the star's interior as a function of $r$. For a static star, the $r$ at which we reach $\rho_{crit}^{mix/pure}$ is the $R^q$, and we get an inner spherical core of mixed-phase inside the HS. For a rotating star the $\rho_{crit}^{mix/pure}$ along different direction ($\theta$) is reached at different $r$. Along 
the equator ($\theta=\frac{\pi}{2}$) it is at ${R_e^q}$ and along the pole ($\theta=0$) it is at ${R_p^q}$, where $R_p^q<R_e^q<R_e$.

The mass fraction (MF) of the quark content inside the star is calculated by taking the ratio of the mass of the quark core with the total mass of the star\\
\begin{align}
	MF=\frac{4\pi}{M}\int_{0}^{R_{e}^q}\int_{0}^{\pi/2}\rho(r,\theta)r^2(\theta)\sin\theta d\theta dr.
	\label{massfraction}
\end{align}
The radial integration can be done up to $R_e^q$ for all values of $\theta$ and is simple for a static star. For a rotating star as $R_p^q<R_e^q$ we do the integration but assume $\rho=0$ for those $\theta$ where $r > R_p^q$. 
The volume for the quark content inside the star is calculated by assuming the star to have an oblate spheroid shape, the volume of such a spheroid is $\frac{4\pi}{3}a^2b$ where $a$ is the equatorial radius and $b$ is the polar radius. Taking the ratio of the volume of the quark core with the volume of the star we define the volume fraction (VF) as
\begin{equation}
	VF=\frac{\left(R_e^q\right)^2R_p^q}{\left(R_e\right)^2R_p}
	\label{volumefraction}
\end{equation}

\section{Results}
\begin{figure}[h]
	\centering
	\includegraphics[width=7.5cm,height=5.5cm]{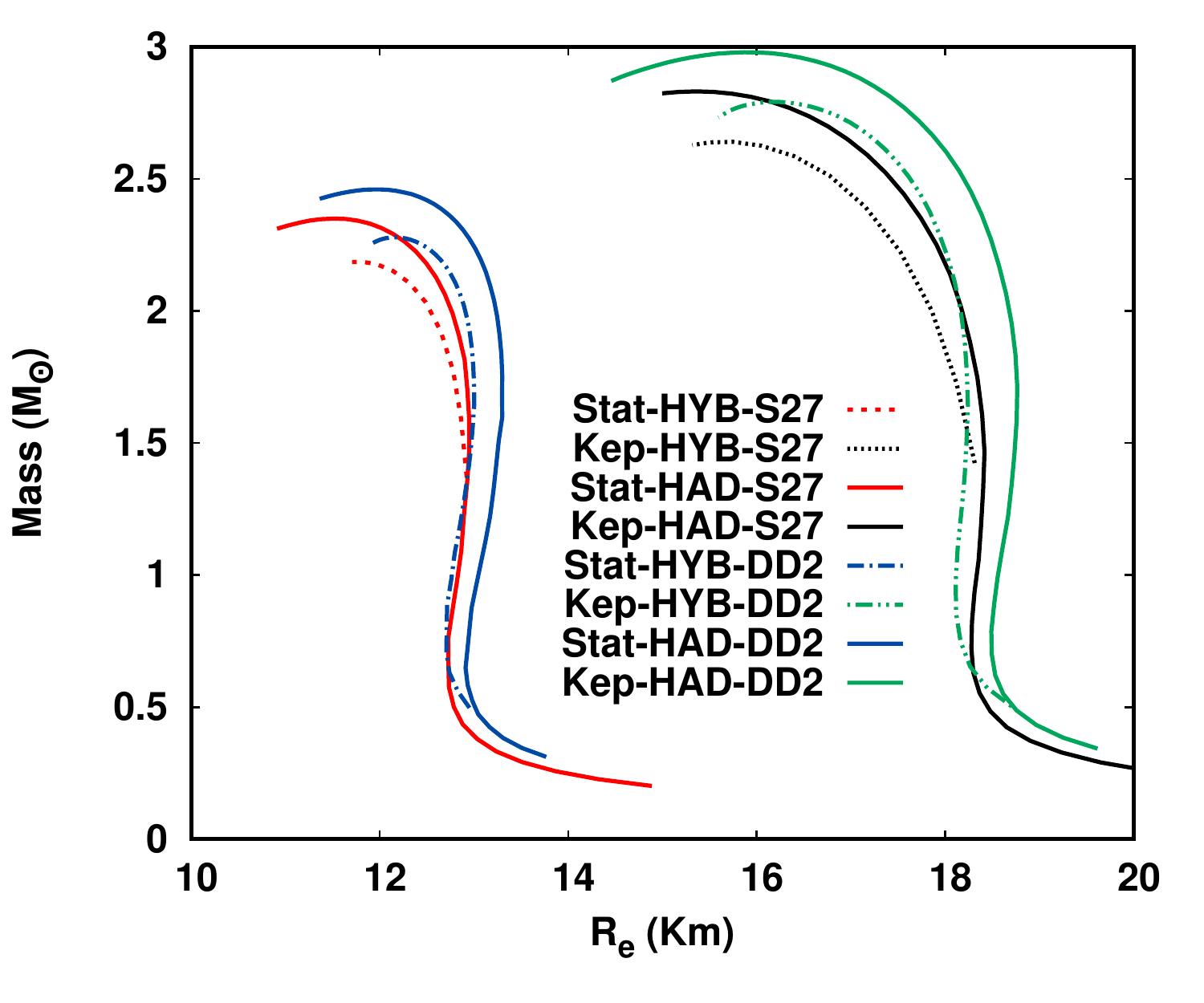}
	\caption{$M-R$ has been plotted for S27 EoS. The plot shows two main clusters, static model (Stat) and keplerian model (Kep), where each of the models has hadronic (HAD) and Hybrid (HYB) EoS. The maximum mass and its corresponding radius and central density are obtained from the plot. The hadronic EoS supports more massive stars than the quark EoS (as the EoS is softer).}	
	\label{MRNL3}
\end{figure}

In figure \ref{MRNL3}, we plot the mass-radius sequence for stars using S27 and BSR EoS. Mass-radius sequences for NSs and QSs (with mixed-phase EoS) are 
shown in the figures.

The $M-R$ curves are clustered into two distinct regions, one for the static stars and the other for keplerian stars. 
The clustering of the curves happens because the keplerian star has the rotational energy to support more massive stars compared to the static star having the same central energy density. Also, due to centrifugal distortion of the keplerian star,  the $R_e$ equatorial radius of the star is greater than that of the static case, and hence the $M-R$ curve shifts to the right as well.\\

The quark EoS are softer in comparison to hadronic EoS 
because of the presence of an extra degree of freedom. 
Thus the hadronic EoS supports more massive stars than the quark EoS, as seen from the $M-R$ curves. 
The maximum mass of each of the EoS can be compared by calculating the ratio $M^k_{max}/M^s_{max}$, which is the ratio of the maximum keplerian mass and the maximum static mass for each EoS. The ratio remains in the range $\sim1.20-1.30$ (see Table II in supplementary materials) across the EoS and is more or less independent of the EoS, 
which is consistent with previous results \cite{breu}. The previous results were done for NSs; however, we find that the range is also valid for HSs. \\

\begin{figure*}[ht!]
	\includegraphics[width=18.5cm,height=4.5cm]{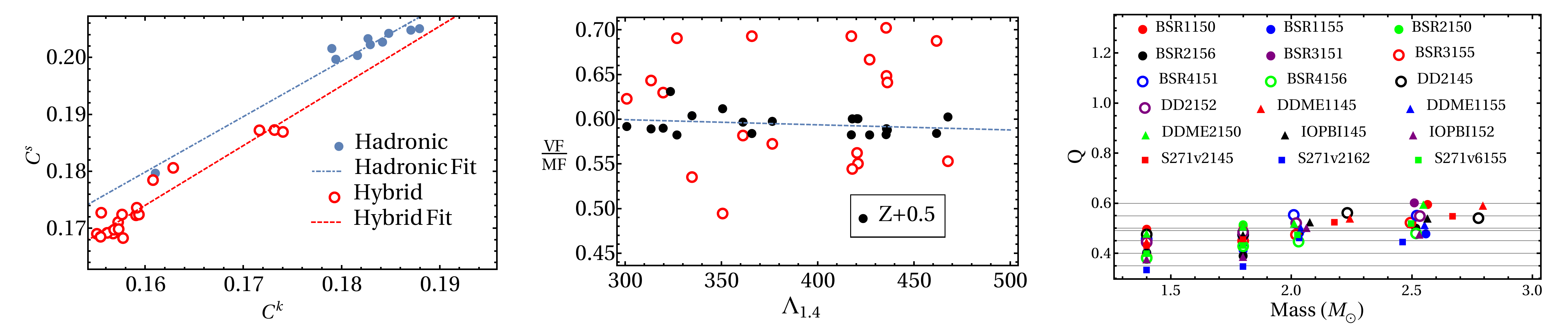}
	\caption{(left panel) Plot of $C^s$ as a function of $C^k$ for NSs and HSs constructed from 20 EoS. The NSs are solid blue dots and the HSs are hollow red dots. (middle panel) Plot of the quark content $VF/MF$ as a function of the tidal deformability $\Lambda_{1.4}$ for $1.4\; M_{\odot}$ static star constructed from 19 different EoS in the red hollow dots. The nonlinear relation of the quark content and the tidal deformation depicts the dependency of both the quark content of the star and the stiffness of the EoS in giving the final tidal deformability value. The  $Z=\left(\frac{MF}{VF}\times\frac{1}{R^k_{1.4}}\right)_{1.4}$ has been plotted along with the quark content and is shifted by $0.5$ for comparison. The Z shows a constant behavior independent of the $\Lambda_{1.4}$ value. (right panel) Plot of Q (the final point of each sequence is $Q^{\#}$) as a function of the mass has been shown. The mass is in the units of solar mass $M_{\odot}$ and has been plotted for mass ranging from $1.0\; M_{\odot}$ to $2.79\; M_{\odot}$ constructed by 20 Gibbs modeled quark EoS. The Q, as can be seen, tends to stay in the band shown by the shaded region in the range lying between $\sim 0.35-0.60$ across all the EoSs.}
	\label{eos1}
\end{figure*}

Similarly, $R_e$ of a keplerian star ($\equiv R^k$) with respect to the static star ($\equiv R^s$) could be captured from the expression $(R^k/R^s)_{1.4}/(R^k/R^s)_{max}$ which is the ratio of keplerian over static radius for a $1.4\; M_{\odot}$ star with the radius of the maximum mass star and its value is given by the linear fit $(1.043\pm0.006)$. 
In figure \ref{eos1} (left panel) we have plotted $C^s$ as a function of $C^k$, where $C^s$ and $C^k$ are the compactness for the maximum static mass star and maximum keplerian mass star, respectively. The fit value for NS is $C^s= 0.0240717 + (0.973707\pm 0.0480356)C^k$ and for HS is $C^s= 0.00631191 + (1.04824\pm 0.0778146)C^k$. The slope of the fit for the NSs is almost similar, but the y-intercept of the NSs is greater than that of HSs. The shift in the curves is due to the appearance of quarks in the HS. As quark appears, the EoS becomes more linear, and therefore the y-intercept reduces.
The ratio of $C^s/C^k$ (defined it as C-R) lies between $1.1-1.26$ for NS and between $1.07-1.11$ for HS, which shows that the shift in the $M-R$ curve is a constant between the static and keplerian stars. This is why grouping the keplerian cases and static cases in the $M-R$ curve irrespective of the EoS. Though there is an overall universality in the C-R, the deviation of the C-R for the NS and HS arises due to the appearance of the quark core. 

The slope of the $C^s$ vs. $C^k$ gives the compactness of the stars. The compactness of HS is more than that of NS, as can be seen from figure \ref{eos1} (left panel). For a given $C^s$, a 
lesser $C^k$ means that the star deformation is lesser upon rotation, precisely the case for HSs.   This is also depicted in the M-R curves plotted above. The radius reduction going from NS to HS is much more significant than the mass reduction, thereby making the compactness of HS more prominent than the compactness of NS. 



The tidal deformability \cite{hinderer}
$\Lambda$ is the measure of how much quadrapole moment $Q_{ij}$ is generated on the star in the presence of an external tidal field $\epsilon_{ij}$ through the 
equation $Q_{ij}=-\lambda\epsilon_{ij}$ where $\Lambda=\lambda M^{-5}$. 
The tidal deformability depends on the compactness of the star, which thereby depends on the EoS. As quark matter softens the EoS, the tidal deformability of the star also depends on the amount of quark content at the star's core. However, in this calculation, we have used different quark EoSs to construct HSs; therefore, $\Lambda$ also depends on the EoS of the quark matter. While constructing the mixed EoS, we matched the hadronic and quark EoS with the mixed-phase region, and so the stiffness and the quark fraction for each EoS are different. Therefore the tidal deformation is not a linear function of the quark content as shown by the red hollow circle in figure \ref{eos1} (middle panel).

The quark content can be quantified by the ratio $\frac{VF}{MF}$ and has been plotted as a function of $\Lambda$ for $1.4\; M_{\odot}$ star in figure \ref{eos1} (middle panel). 
As argued earlier, the $\Lambda_{1.4}$ is not a linear function of quark content and is seen from the figure. However, from the results and figures discussed earlier, we find that the stiffer the hadronic EoS, the larger is their quark content for a fixed mass star.  The keplerian radius $R^k_{1.4}$ gives a good measure of the stiffness of the EoS  of the star because it depicts how much the star could be deformed before it starts to shed mass. Thus we construct a relation given by $\left(Z=\left(\frac{MF}{VF}\times\frac{1}{R^k}\right)_{1.4}\right)$ and has been plotted as a function of quark content (VF/MF) with the fit value $(0.117\pm0.020)-0.00006\lambda_{1.4}$ in figure \ref{eos1} (middle panel).

The quantity $MF/VF$ quantifies the quark content, whereas $R^k_{1.4}$ quantifies the extent to a $1.4$ solar mass NS would maximally deform due to rotation. Therefore, the term signifies how the quark content affects the deformability of the star due to rotation. On the other hand, $\Lambda_{1.4}$ signifies the deformability of the NS due to an external tidal field. Both of them depend on the EoS and, thereby, the quark content of the star. Z is almost a constant (see figure \ref{eos1} (middle panel)) with zero correlation for any EoS as they both portray the same information about the EoS.
In the figure, we have shifted it by a constant factor of $0.5$ so that we can compare it with the variation of $\frac{VF}{MF}$.
The nonlinearity between the quark fraction and the softness of the EoS for a $1.4\; M_{\odot}$ star as a function of $\Lambda_{1.4}$ averages out, and Z becomes constant.

The quark content of a static star and keplerian star can be combined in the form of a relation $Q=\frac{(VF/MF)^k}{(VF/MF)^s}$. The Q measures how much quark content is gained or lost by a keplerian star compared to a static star of the same mass. To calculate the bound on Q, we have plotted all the Q across a range of masses from lower mass stars of $1.0\; M_{\odot}$ to the highest keplerian mass across all EoS which is $2.79\; M_{\odot}$ for DD2 EoS. Between this range, it contains all other masses from all the EoSs and has been plotted in figure \ref{eos1} (right panel). The Q's lie in a patch having the bound of $0.35\le Q\le 0.60$ as can be seen from figure \ref{eos1} (right panel). All the EoS follows the tidal deformability bound $\Lambda_{1.4}\le580$; thus, this Q bound will be a good constraints while construction of EoS satisfying recent observational bound. The Q gives the shift of the quark content of a Keplerian star compared to a static star of the same mass, and the range gives that the shift has to lie between these two values.
The quark content of a keplerain star should reduce by an amount bounded by this range for the star to be stable. A HS with a quark core has to lie within the Q bound to be stable and not start shredding mass or collapse into a black hole (BH).

Also, for each equation of state, we have defined $Q^{\#}$ which has the same definition as Q, but instead of taking the ratio between quark content of static and keplerian stars of the same mass, we take the ratio of the quark content between the maximum static mass star and the maximum keplerian mass star constructed from a given EoS. The $Q^{\#}$ has a range of $\sim0.54-0.60$ universally across the EoS (Table IV in supplementary). This gives the shift in the quark content for a static and a keplerian star for a given EoS. The range shows that the shift in the quark content is much more constrained.
This is following the constancy of the C-R, which says that the shift of the M-R curve is constant between keplerian and static stars independent of the EoS and deviates slightly for NSs and HSs (see Table II in supplementary). The above discussion shows a correlation between the relative compactness of the star and its relative quark content, thus abound on the Q indicates a bound the extent of compactness of a star. 


The HS are thought to be formed after a PT from a NS. Assuming that there is no mass loss during a PT, the baryonic mass of a star remains conserved even after PT. However, the star's gravitational mass is the combination of the baryonic mass and the negative binding energy of the star changes along with the radius of the star. 
\begin{figure*}[ht!]
	\includegraphics[width=17.8cm,height=4.6cm]{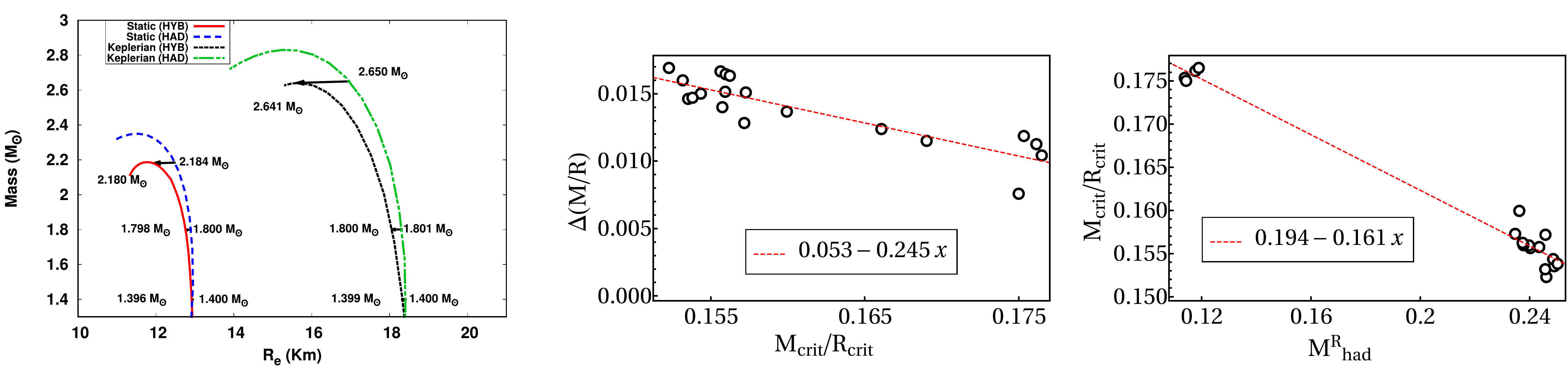}
	\caption{(left panel) plot of gravitational mass as a function of the radius of the star. The plot has been done for static stars constructed form S27 EoS and kinetic stars constructed form S27 EoS. The plot shows PT from NSs to HSs indicated by arrows (keeping the baryonic mass conserved). The mass and radius change can be seen from the plots where the mass loss is most prominent for the highest mass in each case. (middle panel) plot shows the variation of $\Delta(M/R)$ as a function of $M_{crit}/R_{crit}$. (right panel) plot shows the ratio of $M_{crit}/R_{crit}$ as a function of $M^R_{had}$ for different EoS for a PT happening in a NS to a HS with some given value of bag constant and quark coupling strength.The $x$ marked in the figures refer to the $x$ axis of the corresponding figures.}
	\label{mrs}
\end{figure*}

Figure \ref{mrs} (left panel) shows how the gravitational mass and radius of the star changes after PT for S27 EoS.
We see that the change in the gravitational mass is relatively small; however, the radius shrinks considerably. It shows that as PT occurs and a quark core is formed inside a star, the star becomes more compact. We also find that the PT results in exothermic energy generation, and this energy is dissipated in the form of heat, electromagnetic and GWs.
We also find that the mass and radius change due to the PT increases with the star's mass. 
The most interesting observation that can be deduced from the above figure is that massive NS after PT is unstable and most probably collapses to a black hole. This is because the softer EoS (of quark matter) can no longer support such massive stars and, therefore, collapses to a black hole.
There is an upper bound on the NS mass $M_{crit}$ beyond which if an NS undergoes PT becomes unstable and produces a HS not following the Q bound, which either has to shred mass and come under the Q bound or collapse to form a black hole.


An NS is constructed with a given hadronic EoS. For a given hadronic EoS and some given choice of $B_g$ and $a_4$, we can generate a quark EoS satisfying the current bounds. Once this is done, figure \ref{mrs} (right panel) directly gives the critical value of the NS as a function of $M^R_{had}\equiv\frac{M_{had}^{max}(a_4)^4}{\bar{Bag}^{0.05}}$ (where $\bar{Bag}=Bag^{1/4}$) and one can also calculate the maximum mass of the HS it can generate through PT without any baryonic mass loss.
Once we get the critical mass of the NS, we know that more massive NSs up on PT become unstable and collapse to a BH.
Figure \ref{mrs} (middle panel) given a critical value of an NS, we can calculate the change in the compactness of the NS due to a PT, which results in an HS. 

\section{Summary and Conclusion}
Recently, there has been considerable development in astrophysical observation from pulsars and binary NS mergers, which can throw light on the presence of exotic matter at NS cores.
The semi-universal parameters show how the keplerian stars are shifted by a constant amount from their corresponding static counterpart in an $M-R$ sequence and are independent of the EoS. They also show that the deformability of the stars depends both on the quark content and stiffness of the EoS. Also, it is seen that the change in the mass and radius after PT is linearly proportional to the mass of the parent NS and very massive parent NS after PT collapses to a black hole. 
This study also shows how one can find the critical mass of NS and maximum mass of HS for a given EoS, bag constant, and quark coupling constant.

The final discussion is about the change in the properties of an NS to an HS as it undergoes PT.
This change helps deduce the energy output and the GW wave emission when an NS converts to an HS. If the change ratio is small, then the GW emission amplitude is likely to be small, whereas if the ratio of change is high, the amplitude of the GW is likely to be high. Looking at the figures, one can also deduce the GW emission amplitude as a function of the star's mass. If the initial mass is small, the GW emission is small, whereas, for the massive star, the GW emission is significant. For more massive star stable HS configuration is not obtained and the collapses to a BH. 

All the semi-universal relation shows that there is the common bond that all these EoS has to be followed while constructing new EoS. The EoS, though constructed by separate models, satisfies observational bounds and thus has standard features in them in terms of these semi-universal relations. The relations also give an insight into how massive HSs can be and how much quark content can there be in its core without collapsing into a black hole. However, more EoS could be 
tested using these bounds and semi-universal relations to have a better understanding of the interior of such stars.\\

\section{acknowledgement}
DK wishes to acknowledge CSIR India for financial support. The authors are grateful to IISER Bhopal for providing all the research and infrastructure facilities. RM would also like to thank the SERB, Govt. of India, for monetary support in the form of Ramanujan Fellowship (SB/S2/RJN-061/2015).\\

\end{document}


\preprint{APS/123-QED}

\title{Supplemental Material for: Semi Universal relation to understand matter properties at neutron star interiors}

\author{Ritam Mallick}
\email{mallick@iiserb.ac.in}
\affiliation{Department of Physics,\\ Indian Institute of Science Education and Research Bhopal, Bhopal, India.}
\author{Debojoti Kuzur}%
\affiliation{%
	Department of Physics,\\ Indian Institute of Science Education and Research Bhopal, Bhopal, India.
}%
\author{Rana Nandi}%
\affiliation{%
	Department of Physics,Polba Mahavidhyalaya, Hoogly, West Bengal 712148, India.
}%

\date{\today}

\maketitle
\begin{figure*}[h]
	\centering
	\includegraphics[scale=0.65]{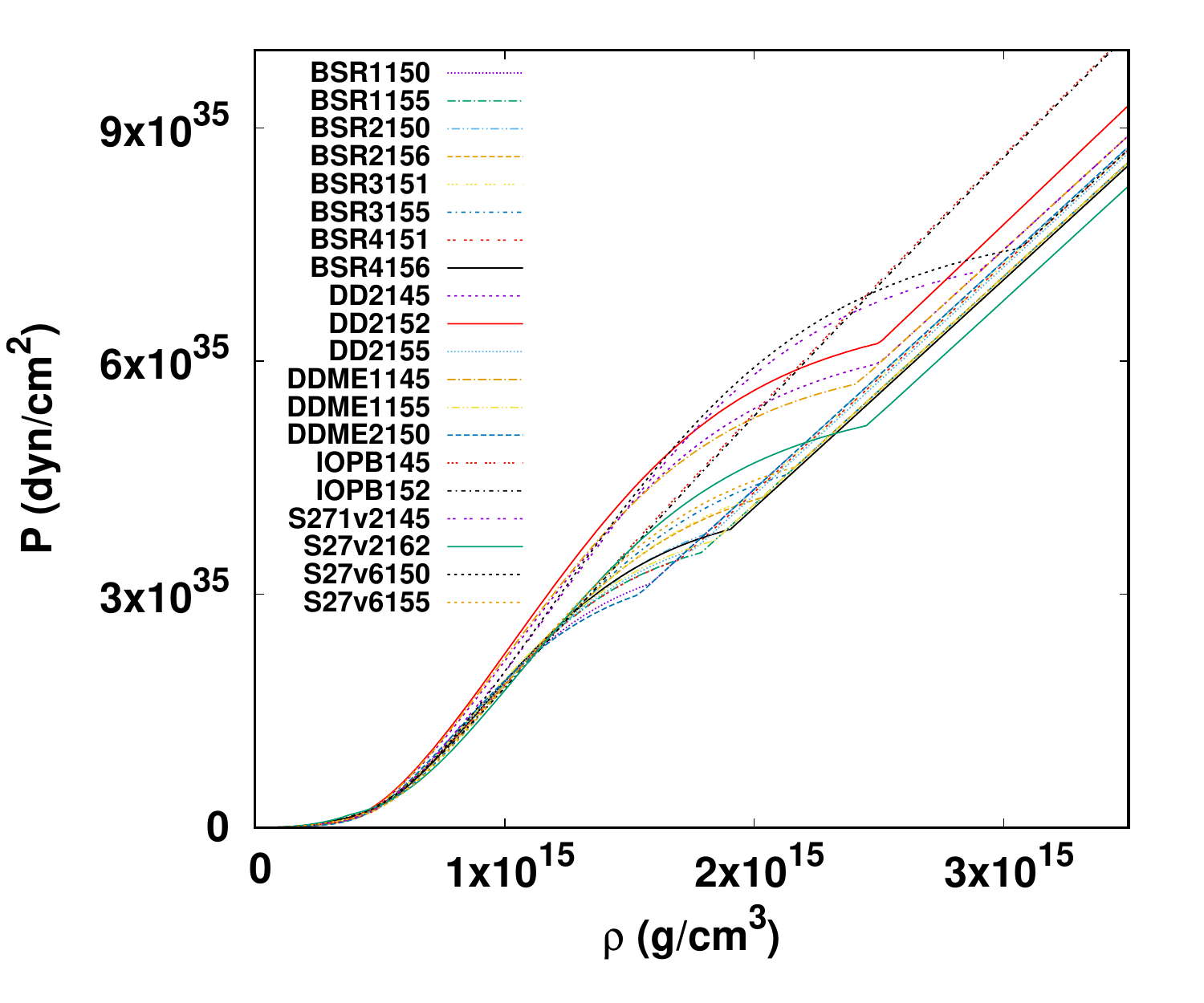}
	\caption{Plot of pressure $P$ as function of the density $\rho$ for 20 EoS using gibbs construction. The change in slope in the plot indicates the presence of mixed phase having quark and HM. The beginning of constant slope shows presence of pure quark matter.}
	\label{eos}
\end{figure*}
\begin{table}[!t]
		\centering
		\caption{Table for densities above which quark matter appears. The critical densities for both mixed and pure quark matter is tabulated for 19 EoSs. The nomenclature of the  EoSs are in the form \textit{hadronic EoS name -bag constant of quark EoS-a4 value of quark EoS} and has been assigned an EoS. No ($N$).}
		\begin{tabular}{|p{1.2cm} |p{3.1cm}||p{3.0cm}|p{3.0cm}|  }
			\hline
			Eos. No ($N$)&EoS &Mixed ($g/cm^3$)&Pure ($g/cm^3$)\\
			\hline
			1&BSR1BPS-150-0.60&$2.21\times 10^{14}$&$1.58\times 10^{15}$\\
			2&BSR1BPS-155-0.60&$3.13\times 10^{14}$&$1.79\times 10^{15}$\\
			3&BSR2BPS-150-0.60&$2.18\times 10^{14}$&$1.80\times 10^{15}$\\
			4&BSR2BPS-156-0.60&$3.28\times 10^{14}$&$2.05\times 10^{15}$\\
			5&BSR3BPS-151-0.60&$2.28\times 10^{14}$&$1.96\times 10^{15}$\\
			6&BSR3BPS-155-0.60&$2.97\times 10^{14}$&$2.13\times 10^{15}$\\
			7&BSR4BPS-151-0.60&$2.26\times 10^{14}$&$1.70\times 10^{15}$\\
			8&BSR4BPS-156-0.60&$3.11\times 10^{14}$&$1.91\times 10^{15}$\\
			9&DD2BPS-145-0.50&$1.48\times 10^{14}$& $2.70\times 10^{15}$\\
			10&DD2BPS-152-0.60&$2.55\times 10^{14}$& $1.76\times 10^{15}$\\
			11&DDME1BPS-145-0.50&$2.32\times 10^{14}$&$2.41\times 10^{15}$\\
			12&DDME1BPS-155-0.60&$3.13\times 10^{14}$&$1.84\times 10^{15}$\\
			13&DDME2BPS-150-0.60&$2.10\times 10^{14}$&$1.54\times 10^{15}$\\
			14&IOPBIBPS-145-0.50&$2.26\times 10^{14}$&$7.34\times 10^{15}$\\
			15&IOPBIBPS-152-0.50&$3.59\times 10^{14}$&$7.79\times 10^{15}$\\
			16&S271v2BPS-145-0.50&$2.20\times 10^{14}$&$2.91\times 10^{15}$\\
			17&S271v2BPS-162-0.60&$3.95\times 10^{14}$&$2.45\times 10^{15}$\\
			18&S271v6BPS-150-0.50&$3.35\times 10^{14}$&$3.06\times 10^{15}$\\
			19&S271v6BPS-155-0.60&$3.08\times 10^{14}$&$2.16\times 10^{15}$\\
			\hline
		\end{tabular}
		\label{table1}
	\end{table}
\begin{table*}[h]
	\caption{Table showing $M^s_{max}$, $M^k_{max}$, $\left(M/R\right)^s_{max}$, $\left(M/R\right)^k_{max}$, $\left(M^k/M^s\right)_{max}$ which are the maximum mass of a static star, maximum mass of a maximally rotating star and their mass by radius raios has been tabulated respectively along with the C-R parameter for 19 EoSs. The tabulation has been done for both hadronic and quark EoS (gibbs construction).}
	\begin{tabular}{|p{1.2cm}||p{8.2cm}|p{8.2cm}|}
		\hline	EoS. No&Hadronic&Hybrid (Gibbs)\\
		\hline
		($N$)&$M^s_{max}$\hspace{0.55cm}$C^s$\hspace{0.85cm}$M^k_{max}$\hspace{0.65cm}$C^k$\hspace{0.95cm}$\left(\frac{M^k}{M^s}\right)_{max}$\hspace{0.35cm}$CR$&$M^s_{max}$\hspace{0.55cm}$C^s$\hspace{0.85cm}$M^k_{max}$\hspace{0.65cm}$C^k$\hspace{0.95cm}$\left(\frac{M^k}{M^s}\right)_{max}$\hspace{0.35cm}$CR$\\
		&&\\
		&($M_{\odot}$)\hspace{1.35cm}\hspace{0.55cm}($M_{\odot}$)\hspace{0.4cm}&($M_{\odot}$)\hspace{1.35cm}\hspace{0.55cm}($M_{\odot}$)\hspace{0.4cm}\\
		\hline
		1&2.469\hspace{0.55cm}0.204\hspace{0.75cm}3.005\hspace{0.65cm}0.185\hspace{0.95cm}1.217\hspace{0.75cm}1.105 &2.011\hspace{0.55cm}0.168\hspace{0.75cm}2.565\hspace{0.65cm}0.158\hspace{0.95cm}1.275\hspace{0.75cm}1.067  \\
		
		2&2.469\hspace{0.55cm}0.204\hspace{0.75cm}3.005\hspace{0.65cm}0.185\hspace{0.95cm}1.217\hspace{0.75cm}1.105 &2.034\hspace{0.55cm}0.169\hspace{0.75cm}2.558\hspace{0.65cm}0.157\hspace{0.95cm}1.258\hspace{0.75cm}1.079  \\
		
		3&2.382\hspace{0.55cm}0.200\hspace{0.75cm}2.886\hspace{0.65cm}0.182\hspace{0.95cm}1.211\hspace{0.75cm}1.103 &2.014\hspace{0.55cm}0.172\hspace{0.75cm}2.534\hspace{0.65cm}0.159\hspace{0.95cm}1.258\hspace{0.75cm}1.083\\
		
		4&2.382\hspace{0.55cm}0.200\hspace{0.75cm}2.886\hspace{0.65cm}0.182\hspace{0.95cm}1.211\hspace{0.75cm}1.103 &2.029\hspace{0.55cm}0.171\hspace{0.75cm}2.516\hspace{0.65cm}0.157\hspace{0.95cm}1.240\hspace{0.75cm}1.088\\
		
		5&2.358\hspace{0.55cm}0.200\hspace{0.75cm}2.851\hspace{0.65cm}0.179\hspace{0.95cm}1.209\hspace{0.75cm}1.113&2.012\hspace{0.55cm}0.172\hspace{0.75cm}2.510\hspace{0.65cm}0.159\hspace{0.95cm}1.247\hspace{0.75cm}1.082\\
		
		6&2.358\hspace{0.55cm}0.200\hspace{0.75cm}2.851\hspace{0.65cm}0.179\hspace{0.95cm}1.209\hspace{0.75cm}1.113&2.019\hspace{0.55cm}0.172\hspace{0.75cm}2.494\hspace{0.65cm}0.158\hspace{0.95cm}1.235\hspace{0.75cm}1.094\\
		
		7&2.441\hspace{0.55cm}0.202\hspace{0.75cm}2.947\hspace{0.65cm}0.183\hspace{0.95cm}1.207\hspace{0.75cm}1.105 &2.010\hspace{0.55cm}0.169\hspace{0.75cm}2.520\hspace{0.65cm}0.156\hspace{0.95cm}1.254\hspace{0.75cm}1.084\\
		
		8&2.441\hspace{0.55cm}0.202\hspace{0.75cm}2.947\hspace{0.65cm}0.183\hspace{0.95cm}1.207\hspace{0.75cm}1.105 &2.030\hspace{0.55cm}0.169\hspace{0.75cm}2.517\hspace{0.65cm}0.155\hspace{0.95cm}1.240\hspace{0.75cm}1.090\\
		
		9&2.416\hspace{0.55cm}0.203\hspace{0.75cm}2.910\hspace{0.65cm}0.184\hspace{0.95cm}1.204\hspace{0.75cm}1.100&2.232\hspace{0.55cm}0.187\hspace{0.75cm}2.776\hspace{0.65cm}0.173\hspace{0.95cm}1.244\hspace{0.75cm}1.081\\
		
		10&2.416\hspace{0.55cm}0.203\hspace{0.75cm}2.910\hspace{0.65cm}0.184\hspace{0.95cm}1.204\hspace{0.75cm}1.100&2.020\hspace{0.55cm}0.170\hspace{0.75cm}2.533\hspace{0.65cm}0.157\hspace{0.95cm}1.254\hspace{0.75cm}1.082\\
		
		11&2.441\hspace{0.55cm}0.205\hspace{0.75cm}2.950\hspace{0.65cm}0.187\hspace{0.95cm}1.208\hspace{0.75cm}1.095 &2.243\hspace{0.55cm}0.187\hspace{0.75cm}2.795\hspace{0.65cm}0.174\hspace{0.95cm}1.246\hspace{0.75cm}1.074 \\
		
		12&2.441\hspace{0.55cm}0.205\hspace{0.75cm}2.950\hspace{0.65cm}0.187\hspace{0.95cm}1.208\hspace{0.75cm}1.095 &2.040\hspace{0.55cm}0.170\hspace{0.75cm}2.553\hspace{0.65cm}0.157\hspace{0.95cm}1.251\hspace{0.75cm}1.080 \\
		
		13&2.481\hspace{0.55cm}0.205\hspace{0.75cm}3.009\hspace{0.65cm}0.188\hspace{0.95cm}1.213\hspace{0.75cm}1.091 &2.012\hspace{0.55cm}0.168\hspace{0.75cm}2.548\hspace{0.65cm}0.155\hspace{0.95cm}1.266\hspace{0.75cm}1.084  \\
		
		14&2.147\hspace{0.55cm}0.180\hspace{0.75cm}2.604\hspace{0.65cm}0.161\hspace{0.95cm}1.213\hspace{0.75cm}1.116 &2.077\hspace{0.55cm}0.181\hspace{0.75cm}2.565\hspace{0.65cm}0.163\hspace{0.95cm}1.235\hspace{0.75cm}1.109  \\
		
		15&2.147\hspace{0.55cm}0.180\hspace{0.75cm}2.604\hspace{0.65cm}0.161\hspace{0.95cm}1.213\hspace{0.75cm}1.116 &2.063\hspace{0.55cm}0.178\hspace{0.75cm}2.531\hspace{0.65cm}0.161\hspace{0.95cm}1.227\hspace{0.75cm}1.110  \\
		
		16&2.336\hspace{0.55cm}0.201\hspace{0.75cm}2.789\hspace{0.65cm}0.179\hspace{0.95cm}1.194\hspace{0.75cm}1.126 &2.179\hspace{0.55cm}0.187\hspace{0.75cm}2.669\hspace{0.65cm}0.172\hspace{0.95cm}1.225\hspace{0.75cm}1.091  \\
		
		17&2.336\hspace{0.55cm}0.201\hspace{0.75cm}2.789\hspace{0.65cm}0.179\hspace{0.95cm}1.194\hspace{0.75cm}1.126 &2.033\hspace{0.55cm}0.173\hspace{0.75cm}2.462\hspace{0.65cm}0.155\hspace{0.95cm}1.211\hspace{0.75cm}1.110  \\
		
		18&2.346\hspace{0.55cm}0.203\hspace{0.75cm}2.819\hspace{0.65cm}0.183\hspace{0.95cm}1.202\hspace{0.75cm}1.112 &2.180\hspace{0.55cm}0.184\hspace{0.75cm}2.640\hspace{0.65cm}0.168\hspace{0.95cm}1.211\hspace{0.75cm}1.102  \\
		
		19&2.346\hspace{0.55cm}0.203\hspace{0.75cm}2.819\hspace{0.65cm}0.183\hspace{0.95cm}1.202\hspace{0.75cm}1.112 &2.026\hspace{0.55cm}0.174\hspace{0.75cm}2.497\hspace{0.65cm}0.159\hspace{0.95cm}1.232\hspace{0.75cm}1.091  \\
		\hline
	\end{tabular}
	\label{table2}
\end{table*}
\begin{table*}[h]
	\centering
	\caption{Table showing $\Lambda_{1.4}$, $(M/R)^s_{1.4}$, $VF^s_{1.4}$ and $MF^s_{1.4}$ which are the tidal deformability, mass by radius ratio, volume fraction and mass fraction for a $1.4\;M_{\odot}$ star respectively along with the Z. The values have been tabulated for 19 EoS for gibbs constructed HSs.}
	\begin{tabular}{|p{1.2cm}||p{9.0cm}|}
		\hline
		\multicolumn{2}{|c|}{Tidal deformability versus Z}\\
		\hline	EoS. No &Hybrid (Gibbs)\\
		\hline
		$(N)$&$\Lambda_{1.4}$\hspace{1cm}$\left(\frac{M}{R}\right)^{s}_{1.4}$\hspace{1.1cm}$VF^s_{1.4}$\hspace{1.25cm}$MF^s_{1.4}$\hspace{1.45cm}$Z$\\
		\hline
		1&435\hspace{1.3cm}0.115\hspace{1.4cm}0.625\hspace{1.4cm}0.890\hspace{1.4cm}0.083\\
		2&376\hspace{1.3cm}0.112\hspace{1.4cm}0.427\hspace{1.4cm}0.746\hspace{1.4cm}0.098\\
		3&365\hspace{1.3cm}0.115\hspace{1.4cm}0.625\hspace{1.4cm}0.902\hspace{1.4cm}0.026\\
		4&334\hspace{1.3cm}0.111\hspace{1.4cm}0.388\hspace{1.4cm}0.725\hspace{1.4cm}0.104\\
		5&435\hspace{1.3cm}0.115\hspace{1.4cm}0.568\hspace{1.4cm}0.876\hspace{1.4cm}0.089\\
		6&361\hspace{1.3cm}0.113\hspace{1.4cm}0.470\hspace{1.4cm}0.808\hspace{1.4cm}0.097\\
		7&313\hspace{1.3cm}0.113\hspace{1.4cm}0.568\hspace{1.4cm}0.883\hspace{1.4cm}0.028\\
		8&420\hspace{1.3cm}0.111\hspace{1.4cm}0.427\hspace{1.4cm}0.776\hspace{1.4cm}0.100\\
		9&319\hspace{1.3cm}0.112\hspace{1.4cm}0.568\hspace{1.4cm}0.902\hspace{1.4cm}0.090\\
		10&300\hspace{1.3cm}0.113\hspace{1.4cm}0.568\hspace{1.4cm}0.912\hspace{1.4cm}0.092\\
		11&326\hspace{1.3cm}0.112\hspace{1.4cm}0.625\hspace{1.4cm}0.905\hspace{1.4cm}0.082\\
		12&420\hspace{1.3cm}0.112\hspace{1.4cm}0.470\hspace{1.4cm}0.836\hspace{1.4cm}0.101\\
		13&461\hspace{1.3cm}0.114\hspace{1.4cm}0.625\hspace{1.4cm}0.909\hspace{1.4cm}0.084\\
		14&417\hspace{1.3cm}0.113\hspace{1.4cm}0.625\hspace{1.4cm}0.902\hspace{1.4cm}0.082\\
		15&350\hspace{1.3cm}0.115\hspace{1.4cm}0.353\hspace{1.4cm}0.714\hspace{1.4cm}0.112\\
		16&436\hspace{1.3cm}0.112\hspace{1.4cm}0.568\hspace{1.4cm}0.886\hspace{1.4cm}0.088\\
		17&323\hspace{1.3cm}0.109\hspace{1.4cm}0.263\hspace{1.4cm}0.646\hspace{1.4cm}0.131\\
		18&418\hspace{1.3cm}0.108\hspace{1.4cm}0.430\hspace{1.4cm}0.790\hspace{1.4cm}0.100\\
		19&467\hspace{1.3cm}0.113\hspace{1.4cm}0.470\hspace{1.4cm}0.850\hspace{1.4cm}0.102\\
		\hline
	\end{tabular}
	\label{table3}
\end{table*}

\begin{table*}[h]
	\centering
	\caption{Table showing mass along with $\Omega$, $VF$, $MF$, and $VF/MF$ of the star respectively. The Q has also been tabulated along with this. The Tabulation is done for 19 gibbs constructed quark EoS. At the end of each EoS, the maximum keplerian mass along with the maximum keplerian frequency has also been tabulated and s indicated by $Q^{\#}$.}
	\begin{tabular}{|p{1.2cm}||p{11.5cm}|}
		\hline	EoS. No & Values\\
		\hline
		$(N)$&$Mass$\hspace{1.55cm}$\Omega$\hspace{1.55cm}$VF$\hspace{1.45cm}$MF$\hspace{1.1cm}$\frac{VF}{MF}$\hspace{1.85cm}$Q$\\
		&\\
		&($M_{\odot}$)\hspace{0.6cm}($\times10^4$ $rad.s^{-1}$)\hspace{5.7cm}$\left[\equiv\frac{(VF/MF)^k}{(VF/MF)^s}\right]$ \\
		\hline
		&1.400\hspace{1.4cm}$0.000$\hspace{1.3cm}0.625\hspace{1.2cm}0.890\hspace{1.0cm}0.702\\
		
		&\hspace{2.13cm}$0.617$ (k)\hspace{0.78cm}0.320\hspace{1.2cm}0.919\hspace{1.0cm}0.348\hspace{1.45cm}0.496\\
		&1.800\hspace{1.4cm}$0.000$\hspace{1.3cm}0.686\hspace{1.2cm}0.918\hspace{1.0cm}0.747\\
		
		1&\hspace{2.13cm}$0.684$ (k)\hspace{0.78cm}0.353\hspace{1.2cm}0.929\hspace{1.0cm}0.380\hspace{1.45cm}0.508\\
		
		\cline{2-2}
		
		&2.011(s-mass)\hspace{0.25cm}$0.000$\hspace{1.3cm}0.754\hspace{1.2cm}0.951\hspace{1.0cm}0.792\\
		
		&\hspace{2.13cm}$0.684$ (k)\hspace{0.78cm}0.388\hspace{1.2cm}0.897\hspace{1.0cm}0.432\hspace{1.45cm}0.545\\
		
		&2.565(k-mass)\hspace{0.21cm}$0.882$\hspace{1.3cm}0.470\hspace{1.2cm}0.997\hspace{1.0cm}0.471\hspace{1.45cm}$0.594^{\#}$\\
		\hline
		&1.400\hspace{1.4cm}$0.000$\hspace{1.3cm}0.427\hspace{1.2cm}0.746\hspace{1.0cm}0.572\\
		
		&\hspace{2.13cm}$0.581$ (k)\hspace{0.78cm}0.159\hspace{1.2cm}0.728\hspace{1.0cm}0.218\hspace{1.45cm}0.381\\
		&1.800\hspace{1.4cm}$0.000$\hspace{1.3cm}0.568\hspace{1.2cm}0.854\hspace{1.0cm}0.665\\
		
		2&\hspace{2.13cm}$0.655$ (k)\hspace{0.78cm}0.216\hspace{1.2cm}0.788\hspace{1.0cm}0.274\hspace{1.45cm}0.412\\
		
		\cline{2-2}
		
		&2.034(s-mass)\hspace{0.25cm}$0.000$\hspace{1.3cm}0.625\hspace{1.2cm}0.894\hspace{1.0cm}0.699\\
		
		&\hspace{2.13cm}$0.701$ (k)\hspace{0.78cm}0.263\hspace{1.2cm}0.782\hspace{1.0cm}0.336\hspace{1.45cm}0.481\\
		
		&2.558(k-mass)\hspace{0.21cm}$0.873$\hspace{1.3cm}0.320\hspace{1.2cm}0.959\hspace{1.0cm}0.334\hspace{1.45cm}$0.477^{\#}$\\
		\hline
		&1.400\hspace{1.4cm}$0.000$\hspace{1.3cm}0.625\hspace{1.2cm}0.902\hspace{1.0cm}0.693\\
		
		&\hspace{2.13cm}$0.616$ (k)\hspace{0.78cm}0.290\hspace{1.2cm}0.932\hspace{1.0cm}0.311\hspace{1.45cm}0.449\\
		&1.800\hspace{1.4cm}$0.000$\hspace{1.3cm}0.686\hspace{1.2cm}0.957\hspace{1.0cm}0.717\\
		
		3&\hspace{2.13cm}$0.689$ (k)\hspace{0.78cm}0.353\hspace{1.2cm}0.959\hspace{1.0cm}0.368\hspace{1.45cm}0.513\\
		
		\cline{2-2}
		
		&2.014(s-mass)\hspace{0.25cm}$0.000$\hspace{1.3cm}0.754\hspace{1.2cm}0.956\hspace{1.0cm}0.789\\
		
		&\hspace{2.13cm}$0.729$ (k)\hspace{0.78cm}0.388\hspace{1.2cm}0.909\hspace{1.0cm}0.427\hspace{1.45cm}0.541\\
		
		&2.534(k-mass)\hspace{0.21cm}$0.901$\hspace{1.3cm}0.427\hspace{1.2cm}0.994\hspace{1.0cm}0.429\hspace{1.45cm}$0.545^{\#}$\\
		\hline
		&1.400\hspace{1.4cm}$0.000$\hspace{1.3cm}0.388\hspace{1.2cm}0.725\hspace{1.0cm}0.535\\
		
		&\hspace{2.13cm}$0.575$ (k)\hspace{0.78cm}0.144\hspace{1.2cm}0.669\hspace{1.0cm}0.215\hspace{1.45cm}0.402\\
		&1.800\hspace{1.4cm}$0.000$\hspace{1.3cm}0.517\hspace{1.2cm}0.828\hspace{1.0cm}0.624\\
		
		4&\hspace{2.13cm}$0.656$ (k)\hspace{0.78cm}0.195\hspace{1.2cm}0.802\hspace{1.0cm}0.243\hspace{1.45cm}0.389\\
		
		\cline{2-2}
		
		&2.029(s-mass)\hspace{0.25cm}$0.000$\hspace{1.3cm}0.568\hspace{1.2cm}0.868\hspace{1.0cm}0.654\\
		
		&\hspace{2.13cm}$0.706$ (k)\hspace{0.78cm}0.238\hspace{1.2cm}0.762\hspace{1.0cm}0.312\hspace{1.45cm}0.477\\
		
		&2.516(k-mass)\hspace{0.21cm}$0.890$\hspace{1.3cm}0.320\hspace{1.2cm}0.978\hspace{1.0cm}0.327\hspace{1.45cm}$0.500^{\#}$\\
		\hline
		&1.400\hspace{1.4cm}$0.000$\hspace{1.3cm}0.568\hspace{1.2cm}0.876\hspace{1.0cm}0.648\\
		
		&\hspace{2.13cm}$0.611$ (k)\hspace{0.78cm}0.263\hspace{1.2cm}0.915\hspace{1.0cm}0.287\hspace{1.45cm}0.443\\
		&1.800\hspace{1.4cm}$0.000$\hspace{1.3cm}0.686\hspace{1.2cm}0.961\hspace{1.0cm}0.714\\
		
		5&\hspace{2.13cm}$0.688$ (k)\hspace{0.78cm}0.320\hspace{1.2cm}0.944\hspace{1.0cm}0.339\hspace{1.45cm}0.475\\
		
		\cline{2-2}
		
		&2.012(s-mass)\hspace{0.25cm}$0.000$\hspace{1.3cm}0.686\hspace{1.2cm}0.961\hspace{1.0cm}0.714\\
		
		&\hspace{2.13cm}$0.731$ (k)\hspace{0.78cm}0.353\hspace{1.2cm}0.891\hspace{1.0cm}0.396\hspace{1.45cm}0.555\\
		
		&2.510(k-mass)\hspace{0.21cm}$0.910$\hspace{1.3cm}0.427\hspace{1.2cm}0.995\hspace{1.0cm}0.429\hspace{1.45cm}$0.601^{\#}$\\
		\hline
		&1.400\hspace{1.4cm}$0.000$\hspace{1.3cm}0.470\hspace{1.2cm}0.808\hspace{1.0cm}0.582\\
		
		&\hspace{2.13cm}$0.584$ (k)\hspace{0.78cm}0.176\hspace{1.2cm}0.791\hspace{1.0cm}0.222\hspace{1.45cm}0.382\\
		&1.800\hspace{1.4cm}$0.000$\hspace{1.3cm}0.568\hspace{1.2cm}0.903\hspace{1.0cm}0.629\\
		
		6&\hspace{2.13cm}$0.668$ (k)\hspace{0.78cm}0.238\hspace{1.2cm}0.842\hspace{1.0cm}0.283\hspace{1.45cm}0.449\\
		
		\cline{2-2}
		
		&2.019(s-mass)\hspace{0.25cm}$0.000$\hspace{1.3cm}0.625\hspace{1.2cm}0.905\hspace{1.0cm}0.691\\
		
		&\hspace{2.13cm}$0.716$ (k)\hspace{0.78cm}0.263\hspace{1.2cm}0.803\hspace{1.0cm}0.327\hspace{1.45cm}0.474\\
		
		&2.494(k-mass)\hspace{0.21cm}$0.900$\hspace{1.3cm}0.353\hspace{1.2cm}0.979\hspace{1.0cm}0.360\hspace{1.45cm}$0.522^{\#}$\\
		\hline
		&1.400\hspace{1.4cm}$0.000$\hspace{1.3cm}0.568\hspace{1.2cm}0.883\hspace{1.0cm}0.643\\
		
		&\hspace{2.13cm}$0.601$ (k)\hspace{0.78cm}0.263\hspace{1.2cm}0.896\hspace{1.0cm}0.293\hspace{1.45cm}0.456\\
		&1.800\hspace{1.4cm}$0.000$\hspace{1.3cm}0.686\hspace{1.2cm}0.961\hspace{1.0cm}0.714\\
		
		7&\hspace{2.13cm}$0.675$ (k)\hspace{0.78cm}0.320\hspace{1.2cm}0.948\hspace{1.0cm}0.337\hspace{1.45cm}0.472\\
		
		\cline{2-2}
		
		&2.010(s-mass)\hspace{0.25cm}$0.000$\hspace{1.3cm}0.686\hspace{1.2cm}0.962\hspace{1.0cm}0.713\\
		
		&\hspace{2.13cm}$0.717$ (k)\hspace{0.78cm}0.353\hspace{1.2cm}0.895\hspace{1.0cm}0.394\hspace{1.45cm}0.553\\
		
		&2.520(k-mass)\hspace{0.21cm}$0.881$\hspace{1.3cm}0.388\hspace{1.2cm}0.989\hspace{1.0cm}0.392\hspace{1.45cm}$0.550^{\#}$\\
		\hline
		&1.400\hspace{1.4cm}$0.000$\hspace{1.3cm}0.427\hspace{1.2cm}0.776\hspace{1.0cm}0.550\\
		
		&\hspace{2.13cm}$0.568$ (k)\hspace{0.78cm}0.144\hspace{1.2cm}0.687\hspace{1.0cm}0.210\hspace{1.45cm}0.381\\
		&1.800\hspace{1.4cm}$0.000$\hspace{1.3cm}0.517\hspace{1.2cm}0.834\hspace{1.0cm}0.620\\
		
		8&\hspace{2.13cm}$0.651$ (k)\hspace{0.78cm}0.216\hspace{1.2cm}0.814\hspace{1.0cm}0.265\hspace{1.45cm}0.428\\
		
		\cline{2-2}
		
		&2.030(s-mass)\hspace{0.25cm}$0.000$\hspace{1.3cm}0.625\hspace{1.2cm}0.906\hspace{1.0cm}0.690\\
		
		&\hspace{2.13cm}$0.700$ (k)\hspace{0.78cm}0.238\hspace{1.2cm}0.773\hspace{1.0cm}0.308\hspace{1.45cm}0.446\\
		
		&2.517(k-mass)\hspace{0.21cm}$0.871$\hspace{1.3cm}0.320\hspace{1.2cm}0.968\hspace{1.0cm}0.330\hspace{1.45cm}$0.479^{\#}$\\
		\hline
	\end{tabular}
	\label{table4}
\end{table*}
\begin{table*}[h]
	\centering
	\begin{tabular}{|p{1.2cm}||p{11.5cm}|}
		\hline
		&1.400\hspace{1.4cm}$0.000$\hspace{1.3cm}0.568\hspace{1.2cm}0.902\hspace{1.0cm}0.629\\
		
		&\hspace{2.13cm}$0.592$ (k)\hspace{0.78cm}0.263\hspace{1.2cm}0.879\hspace{1.0cm}0.299\hspace{1.45cm}0.475\\
		&1.800\hspace{1.4cm}$0.000$\hspace{1.3cm}0.686\hspace{1.2cm}0.956\hspace{1.0cm}0.717\\
		
		9&\hspace{2.13cm}$0.658$ (k)\hspace{0.78cm}0.320\hspace{1.2cm}0.939\hspace{1.0cm}0.341\hspace{1.45cm}0.475\\
		
		\cline{2-2}
		
		&2.232(s-mass)\hspace{0.25cm}$0.000$\hspace{1.3cm}0.754\hspace{1.2cm}0.952\hspace{1.0cm}0.792\\
		
		&\hspace{2.13cm}$0.735$ (k)\hspace{0.78cm}0.388\hspace{1.2cm}0.873\hspace{1.0cm}0.444\hspace{1.45cm}0.561\\
		
		&2.776(k-mass)\hspace{0.21cm}$0.929$\hspace{1.3cm}0.427\hspace{1.2cm}0.998\hspace{1.0cm}0.428\hspace{1.45cm}$0.540^{\#}$\\
		\hline
		&1.400\hspace{1.4cm}$0.000$\hspace{1.3cm}0.568\hspace{1.2cm}0.912\hspace{1.0cm}0.623\\
		
		&\hspace{2.13cm}$0.599$ (k)\hspace{0.78cm}0.238\hspace{1.2cm}0.861\hspace{1.0cm}0.276\hspace{1.45cm}0.444\\
		&1.800\hspace{1.4cm}$0.000$\hspace{1.3cm}0.625\hspace{1.2cm}0.895\hspace{1.0cm}0.698\\
		
		10&\hspace{2.13cm}$0.673$ (k)\hspace{0.78cm}0.320\hspace{1.2cm}0.947\hspace{1.0cm}0.338\hspace{1.45cm}0.484\\
		
		\cline{2-2}
		
		&2.020(s-mass)\hspace{0.25cm}$0.000$\hspace{1.3cm}0.686\hspace{1.2cm}0.959\hspace{1.0cm}0.715\\
		
		&\hspace{2.13cm}$0.715$ (k)\hspace{0.78cm}0.320\hspace{1.2cm}0.860\hspace{1.0cm}0.372\hspace{1.45cm}0.520\\
		
		&2.533(k-mass)\hspace{0.21cm}$0.882$\hspace{1.3cm}0.388\hspace{1.2cm}0.989\hspace{1.0cm}0.392\hspace{1.45cm}$0.548^{\#}$\\
		\hline
		&1.400\hspace{1.4cm}$0.000$\hspace{1.3cm}0.625\hspace{1.2cm}0.905\hspace{1.0cm}0.691\\
		
		&\hspace{2.13cm}$0.595$ (k)\hspace{0.78cm}0.290\hspace{1.2cm}0.936\hspace{1.0cm}0.310\hspace{1.45cm}0.449\\
		11&1.800\hspace{1.4cm}$0.000$\hspace{1.3cm}0.686\hspace{1.2cm}0.957\hspace{1.0cm}0.717\\
		
		&\hspace{2.13cm}$0.661$ (k)\hspace{0.78cm}0.353\hspace{1.2cm}0.961\hspace{1.0cm}0.367\hspace{1.45cm}0.512\\
		
		\cline{2-2}
		
		&2.243(s-mass)\hspace{0.25cm}$0.000$\hspace{1.3cm}0.754\hspace{1.2cm}0.951\hspace{1.0cm}0.793\\
		
		&\hspace{2.13cm}$0.684$ (k)\hspace{0.78cm}0.388\hspace{1.2cm}0.902\hspace{1.0cm}0.430\hspace{1.45cm}0.542\\
		
		&2.795(k-mass)\hspace{0.21cm}$0.929$\hspace{1.3cm}0.470\hspace{1.2cm}0.998\hspace{1.0cm}0.471\hspace{1.45cm}$0.594^{\#}$\\
		\hline
		&1.400\hspace{1.4cm}$0.000$\hspace{1.3cm}0.470\hspace{1.2cm}0.836\hspace{1.0cm}0.562\\
		
		&\hspace{2.13cm}$0.589$ (k)\hspace{0.78cm}0.195\hspace{1.2cm}0.824\hspace{1.0cm}0.237\hspace{1.45cm}0.421\\
		&1.800\hspace{1.4cm}$0.000$\hspace{1.3cm}0.568\hspace{1.2cm}0.895\hspace{1.0cm}0.635\\
		
		12&\hspace{2.13cm}$0.664$ (k)\hspace{0.78cm}0.238\hspace{1.2cm}0.836\hspace{1.0cm}0.285\hspace{1.45cm}0.448\\
		
		\cline{2-2}
		
		&2.040(s-mass)\hspace{0.25cm}$0.000$\hspace{1.3cm}0.625\hspace{1.2cm}0.895\hspace{1.0cm}0.698\\
		
		&\hspace{2.13cm}$0.710$ (k)\hspace{0.78cm}0.290\hspace{1.2cm}0.827\hspace{1.0cm}0.351\hspace{1.45cm}0.502\\
		
		&2.553(k-mass)\hspace{0.21cm}$0.879$\hspace{1.3cm}0.353\hspace{1.2cm}0.979\hspace{1.0cm}0.360\hspace{1.45cm}$0.516^{\#}$\\
		\hline
		&1.400\hspace{1.4cm}$0.000$\hspace{1.3cm}0.625\hspace{1.2cm}0.909\hspace{1.0cm}0.687\\
		
		&\hspace{2.13cm}$0.609$ (k)\hspace{0.78cm}0.320\hspace{1.2cm}0.964\hspace{1.0cm}0.332\hspace{1.45cm}0.419\\
		&1.800\hspace{1.4cm}$0.000$\hspace{1.3cm}0.686\hspace{1.2cm}0.961\hspace{1.0cm}0.714\\
		
		13&\hspace{2.13cm}$0.677$ (k)\hspace{0.78cm}0.353\hspace{1.2cm}0.967\hspace{1.0cm}0.365\hspace{1.45cm}0.511\\
		
		\cline{2-2}
		
		&2.012(s-mass)\hspace{0.25cm}$0.000$\hspace{1.3cm}0.470\hspace{1.2cm}0.955\hspace{1.0cm}0.789\\
		
		&\hspace{2.13cm}$0.714$ (k)\hspace{0.78cm}0.353\hspace{1.2cm}0.939\hspace{1.0cm}0.413\hspace{1.45cm}0.523\\
		
		&2.548(k-mass)\hspace{0.21cm}$0.867$\hspace{1.3cm}0.470\hspace{1.2cm}0.996\hspace{1.0cm}0.472\hspace{1.45cm}$0.598^{\#}$\\
		\hline
		&1.400\hspace{1.4cm}$0.000$\hspace{1.3cm}0.625\hspace{1.2cm}0.902\hspace{1.0cm}0.693\\
		
		&\hspace{2.13cm}$0.599$ (k)\hspace{0.78cm}0.263\hspace{1.2cm}0.906\hspace{1.0cm}0.290\hspace{1.45cm}0.380\\
		&1.800\hspace{1.4cm}$0.000$\hspace{1.3cm}0.686\hspace{1.2cm}0.959\hspace{1.0cm}0.715\\
		
		14&\hspace{2.13cm}$0.671$ (k)\hspace{0.78cm}0.320\hspace{1.2cm}0.938\hspace{1.0cm}0.341\hspace{1.45cm}0.477\\
		
		\cline{2-2}
		
		&2.077(s-mass)\hspace{0.25cm}$0.000$\hspace{1.3cm}0.754\hspace{1.2cm}0.957\hspace{1.0cm}0.788\\
		
		&\hspace{2.13cm}$0.724$ (k)\hspace{0.78cm}0.353\hspace{1.2cm}0.849\hspace{1.0cm}0.416\hspace{1.45cm}0.528\\
		
		&2.565(k-mass)\hspace{0.21cm}$0.920$\hspace{1.3cm}0.427\hspace{1.2cm}0.998\hspace{1.0cm}0.428\hspace{1.45cm}$0.543^{\#}$\\
		\hline
		&1.400\hspace{1.4cm}$0.000$\hspace{1.3cm}0.353\hspace{1.2cm}0.714\hspace{1.0cm}0.494\\
		
		&\hspace{2.13cm}$0.571$ (k)\hspace{0.78cm}0.116\hspace{1.2cm}0.618\hspace{1.0cm}0.187\hspace{1.45cm}0.380\\
		&1.800\hspace{1.4cm}$0.000$\hspace{1.3cm}0.470\hspace{1.2cm}0.788\hspace{1.0cm}0.596\\
		
		15&\hspace{2.13cm}$0.649$ (k)\hspace{0.78cm}0.176\hspace{1.2cm}0.758\hspace{1.0cm}0.232\hspace{1.45cm}0.389\\
		
		\cline{2-2}
		
		&2.063(s-mass)\hspace{0.25cm}$0.000$\hspace{1.3cm}0.568\hspace{1.2cm}0.902\hspace{1.0cm}0.629\\
		
		&\hspace{2.13cm}$0.703$ (k)\hspace{0.78cm}0.216\hspace{1.2cm}0.679\hspace{1.0cm}0.318\hspace{1.45cm}0.505\\
		
		&2.531(k-mass)\hspace{0.21cm}$0.913$\hspace{1.3cm}0.290\hspace{1.2cm}0.962\hspace{1.0cm}0.301\hspace{1.45cm}$0.478^{\#}$\\
		\hline
		&1.400\hspace{1.4cm}$0.000$\hspace{1.3cm}0.568\hspace{1.2cm}0.886\hspace{1.0cm}0.641\\
		
		&\hspace{2.13cm}$0.585$ (k)\hspace{0.78cm}0.238\hspace{1.2cm}0.867\hspace{1.0cm}0.274\hspace{1.45cm}0.428\\
		&1.800\hspace{1.4cm}$0.000$\hspace{1.3cm}0.625\hspace{1.2cm}0.904\hspace{1.0cm}0.691\\
		
		16&\hspace{2.13cm}$0.662$ (k)\hspace{0.78cm}0.290\hspace{1.2cm}0.927\hspace{1.0cm}0.313\hspace{1.45cm}0.452\\
		
		\cline{2-2}
		
		&2.179(s-mass)\hspace{0.25cm}$0.000$\hspace{1.3cm}0.754\hspace{1.2cm}0.962\hspace{1.0cm}0.784\\
		
		&\hspace{2.13cm}$0.740$ (k)\hspace{0.78cm}0.353\hspace{1.2cm}0.863\hspace{1.0cm}0.409\hspace{1.45cm}0.523\\
		
		&2.669(k-mass)\hspace{0.21cm}$0.951$\hspace{1.3cm}0.427\hspace{1.2cm}0.998\hspace{1.0cm}0.428\hspace{1.45cm}$0.546^{\#}$\\
		\hline
		&1.400\hspace{1.4cm}$0.000$\hspace{1.3cm}0.263\hspace{1.2cm}0.646\hspace{1.0cm}0.407\\
		
		&\hspace{2.13cm}$0.539$ (k)\hspace{0.78cm}0.059\hspace{1.2cm}0.438\hspace{1.0cm}0.135\hspace{1.45cm}0.331\\
		&1.800\hspace{1.4cm}$0.000$\hspace{1.3cm}0.388\hspace{1.2cm}0.746\hspace{1.0cm}0.520\\
		
		17&\hspace{2.13cm}$0.637$ (k)\hspace{0.78cm}0.116\hspace{1.2cm}0.648\hspace{1.0cm}0.179\hspace{1.45cm}0.344\\
		
		\cline{2-2}
		
		&2.033(s-mass)\hspace{0.25cm}$0.000$\hspace{1.3cm}0.470\hspace{1.2cm}0.823\hspace{1.0cm}0.571\\
		
		&\hspace{2.13cm}$0.696$ (k)\hspace{0.78cm}0.159\hspace{1.2cm}0.607\hspace{1.0cm}0.262\hspace{1.45cm}0.459\\
		
		&2.462(k-mass)\hspace{0.21cm}$0.892$\hspace{1.3cm}0.238\hspace{1.2cm}0.941\hspace{1.0cm}0.253\hspace{1.45cm}$0.443^{\#}$\\
		\hline
	\end{tabular}
\end{table*}
\begin{table*}[h]
	\centering
	\begin{tabular}{|p{1.2cm}||p{11.5cm}|}
		\hline
		&1.400\hspace{1.4cm}$0.000$\hspace{1.3cm}0.430\hspace{1.2cm}0.790\hspace{1.0cm}0.544\\
		
		&\hspace{2.13cm}$0.558$ (k)\hspace{0.78cm}0.140\hspace{1.2cm}0.720\hspace{1.0cm}0.194\hspace{1.45cm}0.360\\
		18&1.800\hspace{1.4cm}$0.000$\hspace{1.3cm}0.510\hspace{1.2cm}0.920\hspace{1.0cm}0.554\\
		
		&\hspace{2.13cm}$0.654$ (k)\hspace{0.78cm}0.210\hspace{1.2cm}0.800\hspace{1.0cm}0.262\hspace{1.45cm}0.470\\
		
		\cline{2-2}
		
		&2.180(s-mass)\hspace{0.25cm}$0.000$\hspace{1.3cm}0.470\hspace{1.2cm}0.997\hspace{1.0cm}0.471\\
		
		&\hspace{2.13cm}$0.684$ (k)\hspace{0.78cm}0.353\hspace{1.2cm}0.929\hspace{1.0cm}0.380\hspace{1.45cm}0.508\\
		
		&2.640(k-mass)\hspace{0.21cm}$0.926$\hspace{1.3cm}0.330\hspace{1.2cm}0.950\hspace{1.0cm}0.347\hspace{1.45cm}$0.540^{\#}$\\
		\hline
		&1.400\hspace{1.4cm}$0.000$\hspace{1.3cm}0.470\hspace{1.2cm}0.850\hspace{1.0cm}0.552\\
		
		&\hspace{2.13cm}$0.589$ (k)\hspace{0.78cm}0.176\hspace{1.2cm}0.803\hspace{1.0cm}0.219\hspace{1.45cm}0.396\\
		&1.800\hspace{1.4cm}$0.000$\hspace{1.3cm}0.568\hspace{1.2cm}0.869\hspace{1.0cm}0.654\\
		
		19&\hspace{2.13cm}$0.674$ (k)\hspace{0.78cm}0.238\hspace{1.2cm}0.849\hspace{1.0cm}0.280\hspace{1.45cm}0.429\\
		
		\cline{2-2}
		
		&2.026(s-mass)\hspace{0.25cm}$0.000$\hspace{1.3cm}0.625\hspace{1.2cm}0.905\hspace{1.0cm}0.691\\
		
		&\hspace{2.13cm}$0.725$ (k)\hspace{0.78cm}0.263\hspace{1.2cm}0.808\hspace{1.0cm}0.325\hspace{1.45cm}0.471\\
		
		&2.497(k-mass)\hspace{0.21cm}$0.911$\hspace{1.3cm}0.353\hspace{1.2cm}0.990\hspace{1.0cm}0.356\hspace{1.45cm}$0.516^{\#}$\\
		\hline
	\end{tabular}
\end{table*}